\documentclass[fleqn,usenatbib]{mnras}

\usepackage{newtxtext,newtxmath}

\usepackage{pdflscape}

\usepackage[T1]{fontenc}
\usepackage[table]{xcolor}
\DeclareRobustCommand{\VAN}[3]{#2}
\let\VANthebibliography\thebibliography
\def\thebibliography{\DeclareRobustCommand{\VAN}[3]{##3}\VANthebibliography}

\usepackage{graphicx}	
\usepackage{amsmath}	
\usepackage{soul}
\usepackage{xcolor}

\title[Composition and profiles of 12P/Pons-Brooks]{Coma composition and profiles of comet 12P/Pons-Brooks using long-slit spectroscopy}

\author[L. Ferellec et al.]{
Lea Ferellec$^{1}$\thanks{E-mail: lea.ferellec@ed.ac.uk},
Cyrielle Opitom$^{1}$,
Abbie Donaldson$^{1}$,
Johan P. U. Fynbo$^{2}$,
Rosita Kokotanekova$^{1,3}$,
\newauthor Michael S. P. Kelley$^{4}$, Tim Lister$^{5}$\
\\
$^{1}$Institute for Astronomy, University of Edinburgh, Royal Observatory, Edinburgh, EH9 3HJ, UK\\
$^{2}$Cosmic DAWN Center, Niels Bohr Institute, University of Copenhagen, Jagtvej 155, 2200 Copenhagen N, Denmark\\
$^{3}$Institute of Astronomy and National Astronomical Observatory, Bulgarian Academy of Sciences, 72 Tsarigradsko Shose Blvd.,
Sofia 1784, Bulgaria\\
$^{4}$Department of Astronomy, University of Maryland, College Park, MD 20742, USA\\
$^{5}$Las Cumbres Observatory, 6740 Cortona Drive, Suite 102, Goleta, CA 93117, USA
}

\date{Accepted XXX. Received YYY; in original form ZZZ}

\pubyear{\the\year{}}

\begin{document}
\label{firstpage}
\pagerange{\pageref{firstpage}--\pageref{lastpage}}
\maketitle

\begin{abstract}

Comet 12P/Pons-Brook exhibited multiple large and minor outbursts in 2023 on its way to its 2024 perihelion, as it has done during its previous apparitions. We obtained long-slit optical spectra of the comet in 2023 August and 2023 November with the INT-IDS, and in 2023 December with NOT-ALFOSC. Using a standard Haser model in a 10000km-radius aperture and commonly used empirical parent and daughter scale-lengths, our calculated abundance ratios show a constant "typical" composition throughout the period with a C$_2$/CN ratio of about 90 per cent. Molecular density profiles of different species along the slit show asymmetries between opposite sides of the coma and that C$_2$ seems to behave differently than CN and C$_3$. Comparing the coma profiles to a standard Haser model shows that this model cannot accurately reproduce the shape of the coma, and therefore that the calculated production rates cannot be deemed as accurate. We show that an outburst Haser model is a  {slightly} better match to the C$_3$ and CN profile shapes, but the model still does not explain the shape of the C$_2$ profiles and requires equal parent and daughter scale-lengths. Our results suggest that the coma morphology could be better explained by extended sources, and that the nature of 12P's activity introduces bias in the determination of its composition.

\end{abstract}




\section{Introduction} \label{sec:intro}

Comet 12P/Pons-Brooks is a Halley-type comet discovered in 1812, returning to the inner Solar System every 71 years from just beyond the orbit of Neptune on a high-inclination trajectory.
During its 1883 and 1954 apparitions, 12P exhibited multiple outbursts \citep{1883AN....107..131C,1955MNRAS.115..190P}. Approaching its 2024 April 21 perihelion, 12P was recovered in 2020 July \citep{2020RNAAS...4..101Y} when it was already active at 11au. On 2023 July 20, a large outburst was detected with the comet brightening from magnitude 17 to 12\footnote{First reported by Elek Tamás, 
\href{https://www.minorplanetcenter.net/iau/ECS/MPCArchive/2023/MPC_20230912.pdf}{Minor Planet Circular 164789}}. The time of this outburst was then refined to 2023 July 19.57 \citep{2023ATel16194....1M} with a second minor outburst on 2023 July 20.83  \citep{2023ATel16202....1M}.  Since this event 12P has become a target of great interest, and both large and small outbursts were reported along its journey, such as on 2023 September 4 \citep{2023ATel16229....1U}, on 2023 September 22-25 \citep{2023ATel16254....1K}, on 2023 October 5 \citep{2023ATel16270....1U}, on 2023 November 14 \citep{jehin_trappist_2023-2}, on 2023 December 12-14 \citep{jehin_trappist_2024}  {and} on 2024 February 29 \citep{jehin_trappist_2024-1}.

Cometary outbursts are sudden increases in mass loss \citep{1990QJRAS..31...69H} which are thought to be caused by structural failure \citep[e.g. fragmentation or internal gas reservoirs bursting;][]{2004come.book..301B,2024MNRAS.529.2763M}, surface disruptions  \citep[e.g. cliff-collapse or impacts;][]{2017NatAs...1E..92P,2022SoSyR..56..233G}, or physico-chemical processes \citep[e.g. water-ice crystallisation;][]{1974Natur.250..313P} leading to internal pressure building up \citep{2017MNRAS.469S.606A}.

The recent in-situ study of comet 67P/Churyumov-Gerasimenko allowed extensive comparison between topography and outbursts. For 67P, outbursts have been associated to events like cliff-collapse, and correlated with steep slopes and day-night cycles \citep{2016MNRAS.462S.184V}. 
The frequency of large outbursts during each of 12P's approaches suggest that something about its structure, surface or physico-chemical composition {\color{black}is} causing its behaviour, rather than external impactors \citep{2004AN....325..343G}. Using observations from 2024 February, \cite{knight_rotation_2024} determined a rotation period of $57\pm1$hr, much longer than required for rotation-induced fragmentation assuming a typical shape and density \citep{2003Icar..164..492L,2017MNRAS.471.2974K}. 

Observations of 12P during its 2024 approach might help us better understand its structure, activity and make-up. This paper focuses on the composition of the gas released by 12P and its spatial distribution. We obtained long-slit optical spectra of the comet at four different epochs to study abundances and release mechanisms of small radicals in the coma.

\section{Data and reduction methods} \label{sec:data}

\subsection{Isaac Newton Telescope data} \label{subsec:INTdata}

\begin{figure}
    \centering
    \includegraphics[width=\linewidth]{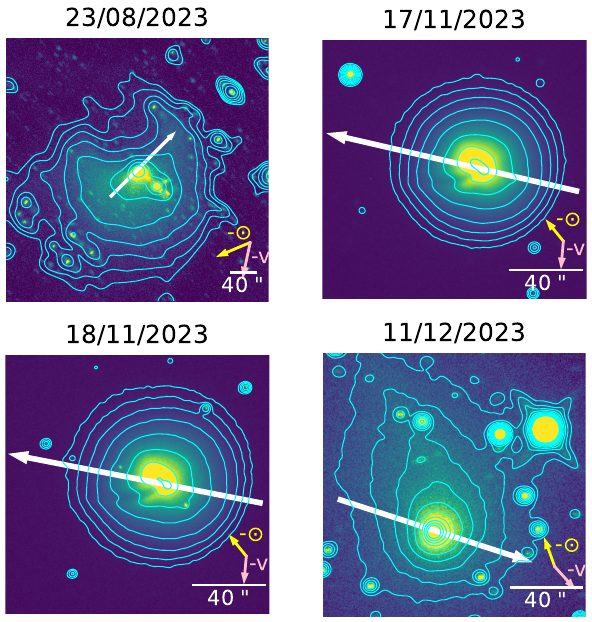}
    \caption{Images of comet 12P from the LCO Outbursting Objects Key Project taken on each observing night, except for 2023 December 17 as the latest available observation was 2023 December 11. It should be noted that an outburst happened on 2023 December 13. On all images North is up and East is left. The slit orientation is represented by a white arrow. The direction of the arrow corresponds to increasing values of the x-axes on Figures \ref{fig:profiles} to \ref{fig:minioutburst}. Blue contours are over-plotted to better highlight the morphology of the dust coma. 
     {Yellow} and pink arrows represent the anti-solar direction and the inverse velocity direction respectively. The LCO data are described in section \ref{subsec:LOOK}.
    }
    \label{fig:look}
\end{figure} 

\begin{figure*}
    \centering
    \includegraphics[width=\linewidth]{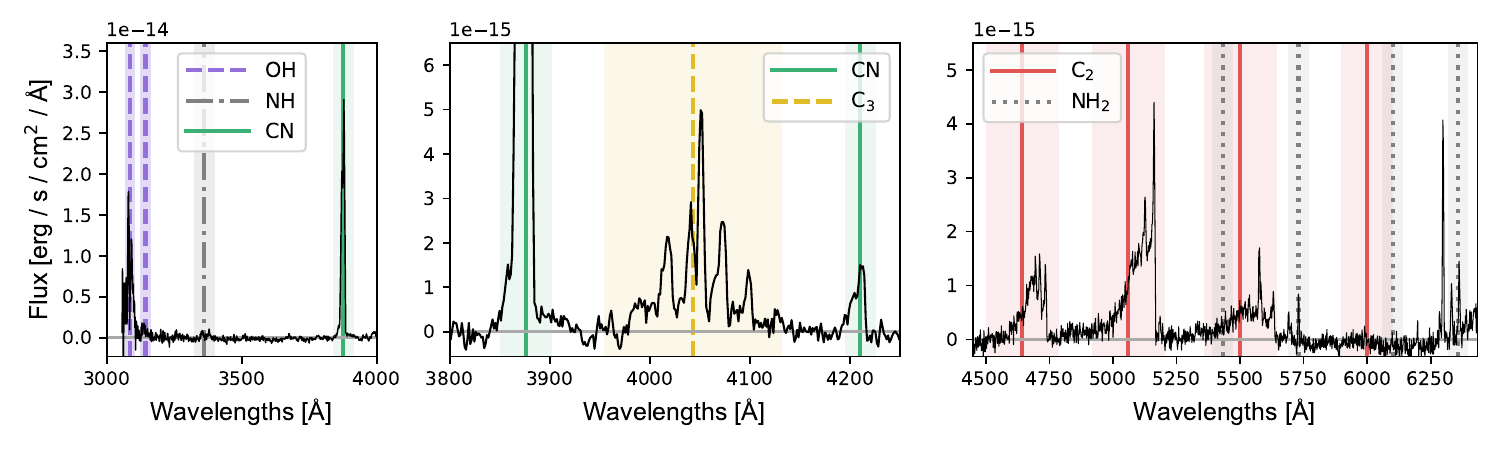}
    \caption{Average INT/IDS dust-subtracted spectrum of 12P within a 10000km radius aperture on 2023 November 17, highlighting the detection of NH, CN, C$_3$, C$_2$ and NH$_2$ emission lines.  {Colored areas denote the approximate extent of each band.} The y-axis scales are different in the three subplots. {Note that there is still evidence of the 5577{\AA} sky emission line.}}
    \label{fig:spectrum}
\end{figure*}

Long-slit spectra of 12P were acquired in 2023 August and 2023 November using the Intermediate Dispersion Spectrograph (IDS) on the INT with the EEV10 detector and a central wavelength of $\sim4500$\AA.  {Grating R400B was used, which has a resolving power of 1596}. With this configuration, the spectra cover the wavelength range of  $\sim3050-6400$\AA, encompassing bright emission lines from OH, NH, CN, C$_2$ and C$_3$ (listed in Table \ref{tab:species_parameters}). The instrument has a total slit length of $3.3$'. A slit-width of $2"$ was used and the instrument has a spatial resolution of $0.4"$/pixel along the slit. 
Three 1200s exposures were acquired on 2023 August 23 23:12 UT {(all times quoted are the mid-times for the observations)}, when the target was at $\Delta\approx3.3$au and $r_{h}\approx3.5$au. Variations in cloud coverage at the time contribute to increased uncertainties on the fluxes measured from this spectrum.
One 900s spectrum and three 600s spectra were obtained consecutively on 2023 November 17 20:31UT, as well as four 600s spectra on 2023 November 18 20:14UT, with clear sky conditions.
The target was at $\Delta\approx2.7$au and $r_{h}\approx2.5$au. 
The slit was aligned with the parallactic angle for all of the spectra. Fig \ref{fig:look} shows images of 12P from the LCO Outbursting Objects Key Project on which we represented our slit orientations. On 2023 November 18, one extra 600s exposure was taken with the slit perpendicular to the parallactic angle to compare coma profiles along different directions, but it was not used for production rate calculations since this orientation is more prone to flux losses from atmospheric refraction, as the INT-IDS does not have an atmospheric dispersion corrector. 

\begin{table}
\centering
\begin{tabular}{lcccc} \hline
Date       & Instrument & Exposures (s)  & $r_h$ (au) & $\Delta$ (au) \\\hline \hline
23/08/2023 & INT-IDS              & $3\times1200$      & 3.5  & 3.3   \\ 
17/11/2023 & INT-IDS              & $1\times900+3\times600$ & 2.5  & 2.7   \\
18/11/2023 & INT-IDS              & $4\times600$$^\dagger$       & 2.5  & 2.7   \\
17/12/2023 & NOT-ALFOSC           & $1\times600$      & 2.2  & 2.4 \\ \hline
\end{tabular}
\caption{ {Summary of our observations (telescope/instrument, number of exposures and exposure time, heliocentric distance $r_h$ and geocentric distance $\Delta$). More details can be found in sections} \ref{subsec:INTdata}  {and} \ref{subsec:NOTdata}.  {$\dagger$: On 2023 November 18, a fifth 600s-exposure spectrum was taken but with a different slit-orientation. It was not used in the final analysis.}}
\label{tab:observations}
\end{table}

\begin{table}
\centering
\begin{tabular}{lcccc}
\hline
Line  & $\lambda$ (\AA)  & $L_{p}/r_{h}^{2}$ [km] & $L_{d}/r_{h}^{2}$ [km] & Ref. \\ \hline \hline
OH(0-0) &  3070-3110  &  $2.40 \times 10^{4} $  & $1.60 \times 10^{5}$   & a              \\ \hline
NH(0-0) &  3320-3400  &  $5.00 \times 10^{4} $  & $1.50 \times 10^{5}$  & b            \\ \hline
CN($\Delta\nu=0$) &  3830-3910  &  $1.30 \times 10^{4} $  & $2.10 \times 10^{5}$ & b\\ \hline
C$_3$ &  3980-4120  &  $2.80 \times 10^{3} $  & $2.70 \times 10^{4}$  & b       \\ \hline
C$_2$($\Delta\nu=0$) &  4860-5180  &  $2.20\times10^{4} $  & $6.60 \times 10^{4}$  & b        \\  \hline
\end{tabular}
\caption{Wavelength ranges used to measure the total flux in prominent emission lines of different species detectable in comet gas spectra, and parent/daughter photo-dissociation scale-lengths $L_p$ and $L_d$ used in the Haser model. References for the scale-lengths: a. \citet{COCHRAN1993}, b. \citet{RANDALL1992}.}
\label{tab:species_parameters}
\end{table}

Wavelength calibration was performed using spectra of ArNe lamps.
The observations of the comet were reduced with bias and flat-field frames taken on each of these nights and were corrected for atmospheric extinction. Atmospheric contamination was removed using spectra of the sky taken $10$' away from the target in between target exposures. Flux calibration was performed using observations of the spectrophotometric standard star BD+284211 compared to reference spectra.
For 2023 November, the flux calibration functions are consistent with our previous INT-IDS observing runs in good weather conditions. Therefore, for each molecular emission region, we estimated the uncertainty associated to flux calibration as the standard deviation of the flux calibration functions between all these observing runs. 
For the data from 2023 August the standard star was only observed the night before observing the target which, combined with variable cloud coverage during these nights, causes a large uncertainty on  {the} flux calibration. However this should affect the overall amplitude of the spectrum more than the relative intensities between lines. For this spectrum we estimated the flux calibration uncertainty as the percentage of difference between the flux calibration function from August and the average flux calibration function from good-weather nights.

The dust contribution to the total comet flux was estimated and removed by adjusting a Sun-like spectrum multiplied by a polynomial slope to the regions in between the expected emission lines. {\color{black}We also excluded both ends of the spectra from the fitting regions because they suffer from increased noise as the illumination of the sensor decreases.} We used spectra of the solar analogue HD186427 acquired during these runs. {\color{black} An order 3 polynomial was used, as it provides a satisfactory match around the emission lines of interest.} We assessed the uncertainty associated to this dust removal process by modifying our dust-model by $\pm5$ per cent and measuring how much this affected the final flux measurements.

\subsection{Nordic Optical Telescope data} \label{subsec:NOTdata}

One 600s spectrum was acquired with the Nordic Optical Telescope on 2023 December 17 at 19:54UT using the ALFOSC instrument. The target was at $\Delta\approx2.4$au and $r_{h}\approx2.2$au.
The spectrum covers a wavelength range of  $\sim$$3500-5350$\AA.
 {Grism 18 has a resolving power of 1000.}
The instrument has a total slit-length of $5.3$'. A slit width of $1"$ was used and the instrument has a spatial resolution of $0.21"$/pixel along the slit. The slit was aligned along the parallactic angle.
Wavelengths calibration was performed using a spectrum of a ThAr lamp. Again, bias and flat-field frames were taken on the same night to reduce the observations of the comet, which were then corrected for atmospheric extinction.

Flux calibration was performed using a spectrum of the spectrophotometric standard star Wolf1346.
This time no offset sky observations were obtained, so gas emissions had to be isolated by adjusting a model of sky and dust together to the continuum. This model was a linear mix of the sky background from the standard star observation and a slope-adjusted composite solar analogue spectrum from our INT runs. 

The target having been observed immediately after the standard star, we assumed that the weather conditions had not changed in between observations, so we used the calibration uncertainties from November as typical variation levels. Continuum removal uncertainties were assessed as described in section \ref{subsec:INTdata}.

\subsection{Las Cumbres Observatory data}
\label{subsec:LOOK}
{\color{black} Images of comet 12P were obtained with the Las Cumbres Observatory (LCO) global telescope network as part of the LCO Outbursting Objects Key Project (LOOK Project; program ID LTP2023B-001) and the Comet Chasers education and public outreach project (program ID FTPEPO2014A-004).  Observations close in time to the spectroscopy were selected: 11 $r'$ images (720 s total) taken 2023 August 23 06:12 UTC from the Faulkes Telescope North (FTN) 2-m at Haleakala Observatory; 1 $R$-band image (60 s) taken 2023 November 17 19:12 from a 1-m telescope at Teide Observatory; 3 $r'$ images (45 s total) taken 2023 November 18 01:02 and 4 $r'$ images (110 s total) taken 2023 December 11 00:54 from 1-m telescopes at McDonald Observatory.  The FTN observations used the MuSCAT3 camera, which simultaneously images through four filters, each filter illuminating a separate 2k$\times$2k CCD with 0.27\arcsec{}~pix$^{-1}$ \citep{narita20}.  The 1-m telescopes used Sinistro cameras, each having a 4k$\times$4k CCD with 0.389\arcsec{}~pix$^{-1}$.  The telescopes followed the comet using the non-sidereal rates from the ephemeris, and the images were combined together by epoch in the rest frame of the comet.  Data were calibrated with LCO's \texttt{BANZAI} data pipeline and photometrically calibrated to the PS1 $r$-band using the \texttt{calviacat} software \citep{mccully18,kelley19-calviacat}. }

\subsection{Molecular production rates calculation} \label{subsec:productionratesmethod}

Spectra were produced by integrating the flux within a $10000$km radius from the nucleus. As an example, our average spectrum from 2023 November 17 is given in Figure \ref{fig:spectrum}.
Total fluxes in emission lines of OH, NH, CN, C$_3$, C$_2$ were calculated by integrating the flux within the ranges given in Table \ref{tab:species_parameters}. 
For emission lines for which the signal is too faint to be detected we calculated 3$\sigma$ upper-limits as described in \cite{cochran_thirty_2012}. These fluxes were converted to total numbers of molecules using fluorescence factors from the \texttt{Lowell Minor Planet Services}\footnote{\hyperlink{https://asteroid.lowell.edu/comet/gfactor}{https://asteroid.lowell.edu/comet/gfactor}}.
Production rates $Q$ of these species were then computed by matching these numbers of molecules to what would be observed in the case of a standard Haser model \citep{1957BSRSL..43..740H}:

\begin{equation}
\centering
    n(r)=\frac{Q}{4 \pi v r^{2}}\frac{L_{d}}{L_{p}-L_{d}}(exp(-\frac{r}{L_{p}})-exp(-\frac{r}{L_{d}}))
    \label{eq:Haser}
\end{equation}

where $n$ is the molecular volume density as a function of cometocentric distance $r$, $L_p$ and $L_d$ the parent and daughter photodissociation scale-lengths, $v$ the expansion velocity of the gas.
This model assumes that the daughter species travel with the same direction and velocity as the parents.  As {in} \cite{1995Icar..118..223A},  {we used the scale-lengths relationships listed in Table }\ref{tab:species_parameters},  {as well as a velocity of $v=1$km~s$^{-1}$ regardless of heliocentric distance. In reality, using a constant unity velocity yields measurements of $Q/v$, but we  will label our results $Q$ from now on.}
The resulting production rates are presented and discussed in section \ref{subsec:productionrates}.  {It should be noted that, if the model and scale-lengths do not accurately represent the coma, then the resulting production rates are non-physical and aperture-dependent. Hence we chose a 10000km aperture to at least guarantee that we can compare our measurements with other published results using this same aperture (see section }\ref{subsec:productionrates} {). The validity of the standard Haser model for 12P is tested in section} \ref{sec:standardhaser}.

\subsection{Radial molecular column density profiles} \label{subsec:radialprofilesmethod}

From the spatial information contained in the long-slit observations, radial profiles of the column density of molecules in the coma were computed for species with the brightest emissions: CN, C$_2$ and C$_3$. We did not compute profiles for OH as the emission line is at the very edge of the wavelength range covered by the instrument, making the dust-subtraction less reliable.

To produce these profiles, we removed the dust contribution from each individual spectrum at each location along the slit, allowing us to measure the fluxes within the gas emission ranges at each location, which we converted into column density versus nucleocentric distance. These profiles are shown and described in section \ref{subsec:radialprofiles}.
By binning the data we created smoothed profiles for visualisation only, but used the full data set for the least-square optimisation of the models described below.

In sections \ref{sec:standardhaser} to \ref{sec:minioutburst}, we compare these profiles to the standard Haser model (eq. \ref{eq:Haser}) as well as to an outburst model from \cite{2016A&A...589A...8O} in which $Q/v$ from the standard Haser model is replaced by the following expression, which accounts for an exponential increase up to the outburst peak then an exponential decrease back to steady-state: 

\begin{equation}
\label{eq:Haseroutburst1}
    Q(r>v_{1}{\Delta}t)=\frac{Q_{0}}{v}+\frac{Q_{1}}{v_{1}}exp(-\frac{v_{1}{\Delta}t-{r}}{r_a})
\end{equation}

\begin{equation}
\label{eq:Haseroutburst2}
    Q(r<v_{1}{\Delta}t)=\frac{Q_{0}}{v}+\frac{Q_{1}}{v_{1}}exp(-\frac{{r}-v_{1}{\Delta}t}{r_b})
\end{equation}

where  $Q_{0}$ is the steady-state production rate, $Q_{1}$ corresponds to the additional outburst gas release {\color{black} (so that at the peak of the outburst the total production rate is $Q_{0}+Q_{1}$)}, $v_1$ the expansion velocity of the outburst material, $r{_a}/v_1$ and $r_{b}/v_1$ the respective characteristic timescales of the increase and decrease in activity during the outburst, and ${\Delta}t$ is the time {\color{black} between the peak of the outburst and the time of observation. When adjusting these models to our data, we will alternatively refer to the best-fit outburst peak date instead of ${\Delta}t$.}

\begin{table*}
    \centering
    \begin{tabular}{lrcccc}
    \cline{3-6}
       &             & 23/08/2023 & 17/11/2023 & 18/11/2023 & 17/12/2023 \\ \hline \hline
    OH  & Q [$10^{28}$~s$^{-1}$]  & $< 3.71$      &$5.71\pm0.60 $   &$5.35\pm0.55$  &        --       \\ \hline
    CN & Q [$10^{26}$~s$^{-1}$]  &$0.218\pm0.110$&$2.10\pm0.13$    &$1.62\pm0.10$  & $1.63\pm0.10$   \\
       & Q/Q$_{OH}$ [\%]  &  --            &$0.367\pm0.044$  &$0.304\pm0.036$&      --      \\ 
    & log(Q/Q$_{OH}$)   &  --            &$-2.44\pm0.05$  &$-2.52\pm0.05$&     --     \\\hline
    
    C$_2$& Q [$10^{26}$~s$^{-1}$]  &  $< 0.436$    &$1.93\pm0.16$    &$1.46\pm0.12$  &  $ 1.50\pm0.12$   \\ 
     & Q/Q$_{OH}$ [\%]  &   --           &$0.338\pm0.045$  &$0.273\pm0.036$&    --        \\ 
       
       & Q/Q$_{CN}$ [\%]  & $< 200$       &$92.2\pm9.3$     &$89.8\pm9.1$  &  $92.0\pm9.4$       \\
        & log(Q/Q$_{OH}$) &   --           &$-2.47\pm0.06$  &$-2.56\pm0.06$&    --        \\
      & log(Q/Q$_{CN}$) & $< 4.30$       &$-0.035\pm0.046$     &$-0.046\pm0.046$  &  $-0.036\pm0.047$       \\\hline
    C$_3$ & Q [$10^{25}$~s$^{-1}$]  &  $<0.193$     &$1.20\pm0.13 $   &$0.905\pm0.081$&  $ 1.00\pm0.10$            \\
       & Q/Q$_{OH}$ [\%]  &  --           &$0.0210\pm0.0031$& $0.0169\pm0.0023$&      --      \\
        
       & Q/Q$_{CN}$ [\%]  &  $<8.85$      &$5.72\pm0.70$    &  $5.56\pm0.60$&   $6.20\pm0.70$     \\
       & log(Q/Q$_{OH}$)  &   --           &$-3.68\pm0.07$& $-3.77\pm0.06$&      --      \\
    & log(Q/Q$_{CN}$)  &  $<1.05$      &$-1.24\pm0.06$    &  $-1.25\pm0.05$&   $-1.20\pm0.05$     \\ \hline
    NH & Q [$10^{26}$~s$^{-1}$]  & $< 4.82$      &$2.22 \pm 0.19 $ &  $< 2.76$     &     --       \\
     
         & Q/Q$_{OH}$ [\%]  &  --            &$0.389\pm0.052$  &   $<0.515$    &      --      \\
       & Q/Q$_{CN}$ [\%]  & $<2208$       &$106\pm11$       &   $<169$      &     --       \\

   & log(Q/Q$_{OH}$)  &  --            &$-2.41\pm0.06$  &   $<-2.28$    &      --      \\

    & log(Q/Q$_{CN}$)  & $<1.34$       &$0.0253\pm0.0475$       &   $<0.228$      &     --       \\\hline
    \end{tabular}
    \caption{Production rates calculated from the total fluxes within a 10000km aperture using a standard Haser model with v=1km~s$^{-1}$, g-factors from the \texttt{Lowell Minor Planet Services} and photodissociation scalelengths from Table \ref{tab:species_parameters}. Upper limits were calculated according to \citep{cochran_thirty_2012}.  {"--"} indicates production rates that could not be measured (because the line is outside of the instrument's coverage) or ratios that could not be calculated (as the denominator is an upper limit).}
    \label{tab:productionrates}
\end{table*}

\section{Results and discussion} \label{sec:resukts}

\subsection{Gas Composition} \label{subsec:productionrates}

The molecular production rates and abundance ratios that we calculated as explained in section \ref{subsec:productionratesmethod} are given in Table \ref{tab:productionrates}. {As a reminder, we are using the emission lines listed in Table} \ref{tab:species_parameters}. Following the 2023 November 14 outburst, we observe a rapid decrease of the production rates for all molecules ($\sim$20-25 per cent for C$_2$, C$_3$, CN and  6 per cent for OH between November 17 and 18). A month later, the outgassing rates are similar to November 18 as the effects of the outburst have cleared but the comet has gotten closer to the Sun.

Our abundance ratios indicate that 12P has a "typical" composition according to the classification made by \cite{1995Icar..118..223A}, as opposed to "C$_2$-depleted" comets. However the C$_2$/CN ratio is below the average for the "typical" class. It should be noted that we use the same scale-length relationships as \cite{1995Icar..118..223A}.
Considering the C$_2$-depletion condition of $Q(C_{2})/Q(CN) < 77$ per cent revised by \cite{2003Icar..162..415S}, our measurements show that 12P is close to the depletion limit, with an average C$_2$/CN ratio of $91.3\pm5.3$ per cent in November-December.  We do not detect any significant composition change while the outburst settles nor between November and December.\\

\begin{figure}
    \centering
    \includegraphics[width=\linewidth]{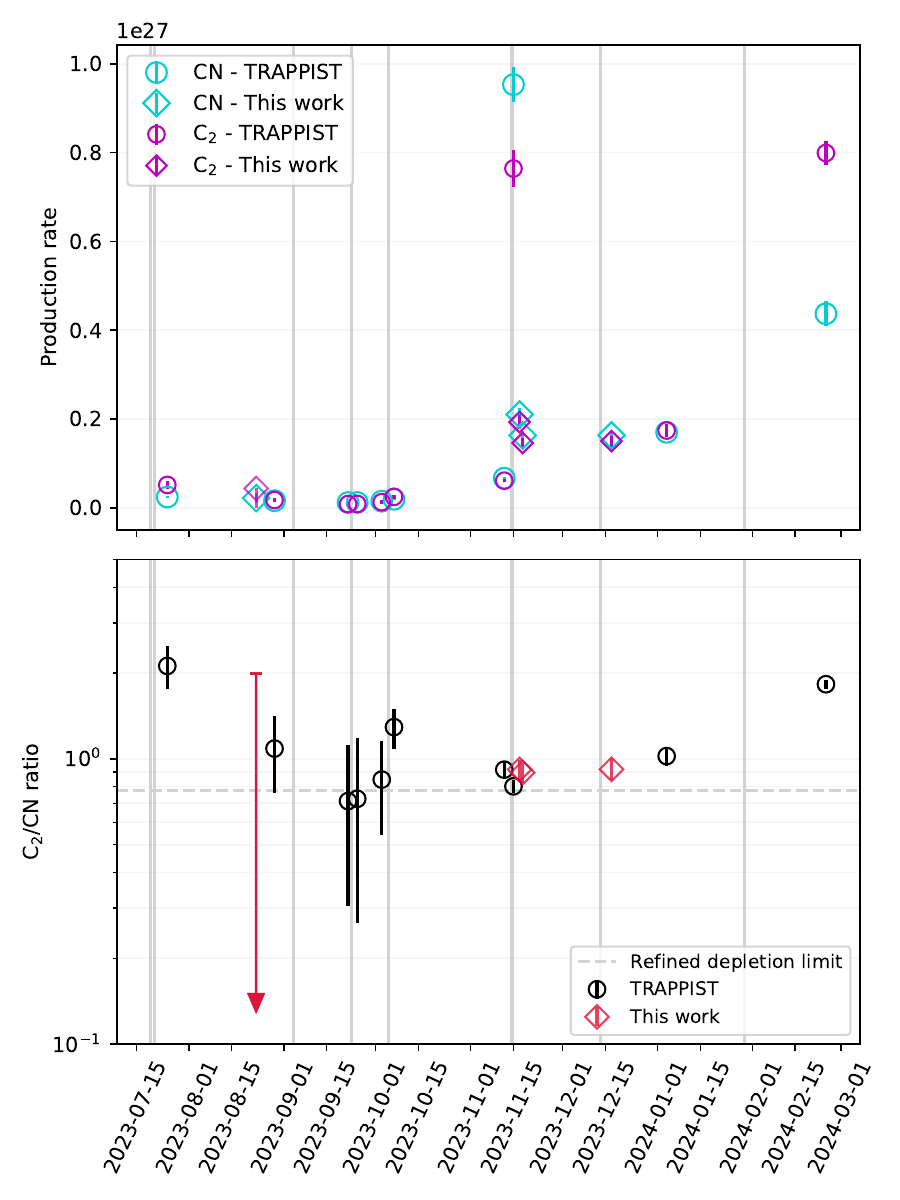}
    \caption{Production rates of CN and C$_2$ (top panel) and C$_2$/CN abundance ratio (bottom panel) between 2023 July and 2024 March from our measurements  {(diamond markers)} and preliminary values from TRAPPIST \citep{jehin_trappist_2023, jehin_trappist_2023-1, jehin_trappist_2023-2, jehin_trappist_2024, jehin_trappist_2024-1}  {(circular markers)}. The arrow on the bottom panel represents an upper limit, as our C$_2$ production rate for 2023 August 23 is an upper limit. Vertical lines indicate reported outbursts. On the bottom panel, a dashed horizontal line indicates the carbon-depletion threshold from the survey by \citet{1995Icar..118..223A} updated by \citet{2003Icar..162..415S}.}
    \label{fig:trappist}
\end{figure}

Our results are also consistent with the preliminary production rates published by \cite{jehin_trappist_2023, jehin_trappist_2023-1, jehin_trappist_2023-2, jehin_trappist_2024, jehin_trappist_2024-1} using observations from the TRAPPIST telescopes. While their production rates derive from narrowband-filter images, most of them were calculated using the same parameters and aperture radius as our analysis (except for their measurements from September 25, September 27, October 03 and October 07 which use v=0.5km~s$^{-1}$ and a 100000km aperture radius). These measurements along with ours are represented in Figure \ref{fig:trappist}, and show an overall increase of the production rates as the comet approaches the Sun, as well as an increase by more than a factor 10 throughout the November 14 outburst{\color{black}. The outburst then seems to rapidly settle.} 
The TRAPPIST observations yield C$_2$/CN ratios similar to our findings ($91.7\pm6.0$ per cent on November 12 and $80.2\pm4.3$  per cent on November 15) but show that the ratio varies greatly across the semester (see Figure \ref{fig:trappist}). While variations of C$_2$/CN ratios with heliocentric distance have been reported by multiple studies \citep[e.g. ][]{1995Icar..118..223A,2011Icar..213..280L}, the behaviour observed in 12P seems to differ from these trends which would predict a smooth increase throughout the comet's approach and do not explain the decrease observed from July to September.

Other studies have measured how outbursts affect the apparent composition of a comet, or how homogeneous the interior of comets seem to be.
\cite{2003Icar..162..415S} measured the same composition in multiple fragments of comet 73P/Schwassmann–Wachmann as before it fragmented.
\cite{2008ApJ...680..793D} studied comet 17P after a major outburst, which did not seem fragmentation-related, and also did not find any notable composition change.
\cite{ORBi-020422a4-b0c0-4088-b20f-94f93cd7a95c} measured production rates of comets 168P/Hergenrother, C/2010 G2 (Hill), C/2012 S1 (Ison), and C/2013 A1 (Siding Spring) throughout outbursts, and while different mechanisms are believed to be the cause of these outbursts, they also did not detect composition changes. 
These studies suggest that these comets have a homogeneous composition, throughout their whole interior or at least in the outer layers.
Our measurements along with the TRAPPIST ones also show no significant composition change through the November 14 outburst, however \cite{jehin_trappist_2023-1,jehin_trappist_2023-2} measure an increase of C$_2$/CN after the October outburst. 
It could be that this is representative of the outburst material, however we will illustrate in section \ref{sec:standardhaser} how the {\color{black} atypical} coma shape of 12P can introduce bias in the measurements of its composition.\\

\subsection{Dust spectrum properties}
\label{subsec:dust spectrum}

{\color{black}
For the data from 2023 November, we computed the reflectance of the coma by dividing the non-dust-removed comet spectrum by the solar analogue spectrum. We calculated the dust reflectance slope by normalizing the reflectance at 5200\AA\, then performing a linear regression on the data restricted to the ranges 4400-4500\AA, 4750-4900\AA, 5200-5320\AA\, and 6100-6200\AA\, as to avoid gas emission lines. Lower wavelengths were not included as the reflectance was noisier. We did not compute dust spectral slopes for the data from 2023 August as  {it has} a significantly lower signal-to-noise ratio and suffers from bad weather conditions, nor for the data from 2023 December as the dust and sky contributions cannot be reliably separated. 

We obtain a spectral slope of $(4.5\pm0.3)\%$/1000\AA\, on average for 2023 November 17-18 between 4400 and 6200\AA, where the uncertainty reflects the error on the mean among the 8 spectra considered. This is lower than the slopes of 15$\%$/1000\AA\,  to 39$\%$/1000\AA\, measured by \cite{1992Icar...98..163S} in the 4400-5675\AA\, range among 18 comets, or the average of $(13\pm5)\%$/1000\AA\, measured by \cite{1986ApJ...310..937J} over 3500-6500\AA\, (9 comets).

However we do see a steeper reflectance in the bluer end of our range. Using the 4400-4500\AA\, and 4750-4900\AA\, regions only we measure an average slope of $(10.8\pm0.6)\%$/1000\AA\, (this time with the reflectance normalized at 4760\AA). This is comparable to some of the spectral slopes measured by \cite{ee7de70a362348b18ae3b3c84f690742} between 4450 and 5260\AA, with $(13\pm8)\%$/1000\AA\, on average. As for other studies of this type, we find that the reddening of the reflectance gets lesser at higher wavelengths.}

\subsection{Coma profiles} \label{subsec:radialprofiles}

\begin{figure*}
    \centering
    \includegraphics[width=\linewidth]{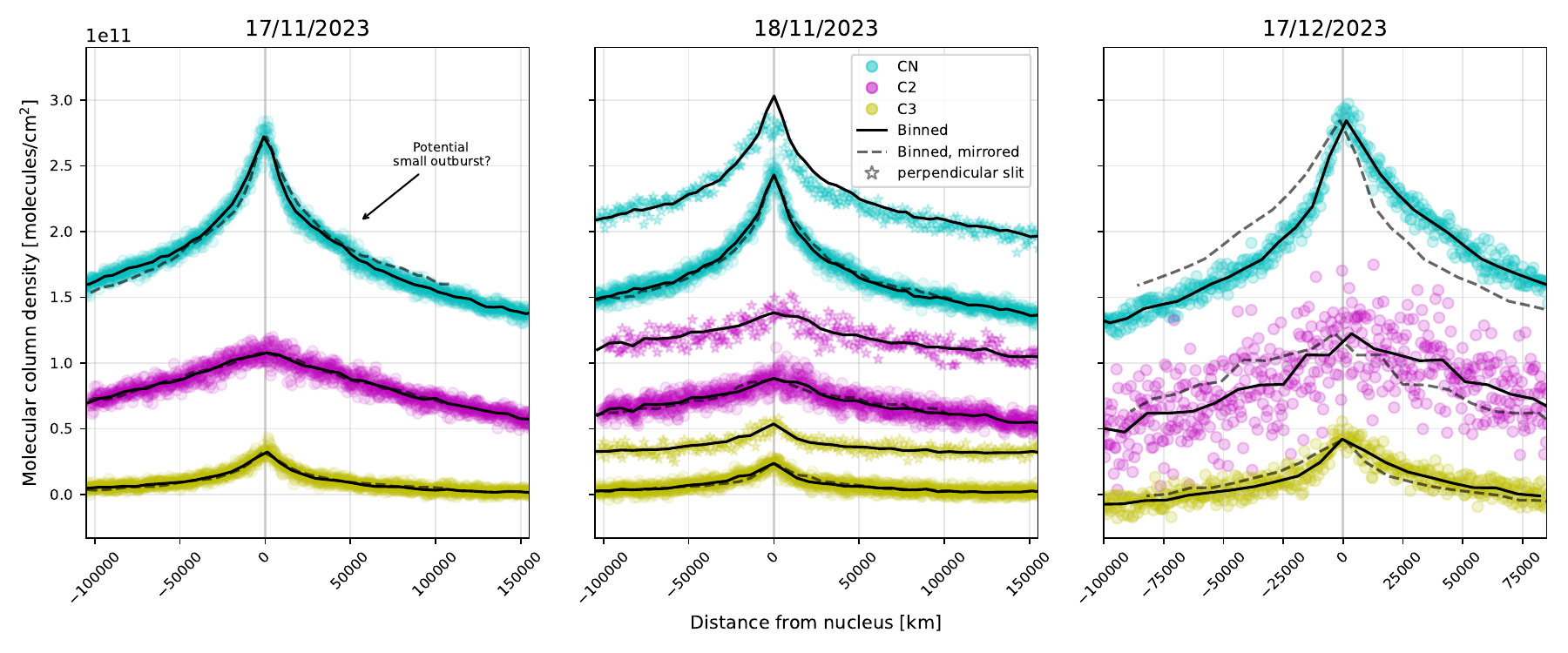}
    \caption{Density profiles of CN (cyan), C$_2$ (magenta) and C$_3$ ( {yellow}) versus nucleocentric distance on 2023 November 17 (left), 2023 November 18 (middle) and 2023 December 17 (right). Circular markers represent profiles along the slit oriented along the white arrows shown on Figure \ref{fig:look}, with the arrow pointing towards positive distance values on this figure. For 2023 November 18, profiles with star-shaped markers are from a single exposure with the slit perpendicular to the orientation shown by the white arrow.
    The following offsets were added to allow for easier visualisation: for circular markers $1\times10^{11}$ for CN and  $0.2\times10^{11}$ for C$_2$, for star-shaped markers  $1.6\times10^{11}$ for CN, $0.7\times10^{11}$ for C$_2$ and $0.3\times10^{11}$ for C$_3$. Solid black lines are smoothed (binned) profiles made from the profiles aligned with the parallactic angle (even those plotted over the perpendicular profiles, to allow comparison between both orientations). Dashed grey lines are these same binned profiles reversed along the x-axis to highlight the asymmetry in the coma.}
    \label{fig:profiles}
\end{figure*}

\subsubsection{General aspect}

Column density profiles for CN, C$_2$ and C$_3$ are shown on Figure \ref{fig:profiles}{, using the same emission lines that were used for the composition measurements.}  Figure \ref{fig:standardhaser} {presents these same profiles plotted with logarithmc scales. This makes it clear that the inner parts of the profiles are relatively
flat. Some of this flatness could be the result of seeing or guiding errors.} Figure \ref{fig:look} shows the slit orientation with respect to the observed coma morphology around the time of observation, for comparison between our gas profiles and the apparent dust distribution. Note that for December the image was taken a few days before our observations, and a small outburst was reported in between.

\paragraph*{Asymmetry:} 

 {Unlike the dust, pushed tailward by solar radiation, imaging has revealed that comets can exhibit diverse gas coma morphologies such as fans or jets} \citep[e.g.][]{2021PSJ.....2..104K}.  {Depending on the origin of the gas species and outgassing behaviour of the comet, the orientations of these features are not necessarily linked to the direction of the Sun. They may instead be dictated by the rotation state of the comet, and might vary between species. While we cannot obtain a full picture of the gas coma, we can look into the symmetry of our observed density profiles along the slit.}

On November 17 and November 18, the profiles present a slight asymmetry, with the left side  of the profiles (South-West of the coma) showing a slightly higher density of molecules than the right side (North-East of the coma). This is mostly visible for the CN profiles which have a higher signal-to-noise ratio than the other species, and the asymmetry is more {\color{black} pronounced} on November 17 than on November 18, which could indicate that is it linked to the outburst material. Figure \ref{fig:look}, as well as our own analysis of the dust component in our spectra, indicate that the North-East side of the coma seems to carry significantly more dust at the time. 
{\color{black} This orientation coincides with the anti-Solar direction expected for a dust-tail. Therefore this dust distribution does not necessarily represent any preferred direction in the steady-state or outburst dust ejection, making it difficult to correlate with the gas distribution.}

On November 18th, the profiles acquired perpendicularly to the parallactic angle do not show any significant difference with the the ones aligned with the parallactic angle for C$_2$ and C$_3$, but for CN the peak of the profile at the nucleus seems less sharp than for the initial orientation.

On December 17, the asymmetry between both sides is even stronger, with higher molecular densities on the right side (South-West) than the left side (North-East), suggesting anisotropic gas release.  \cite{knight_rotation_2024} reported a complex coma morphology in 2024 February, displaying rotating CN jets that could explain the asymmetry that we observe, although the jets were 180 degrees apart in their viewing geometry. The image of the coma shown on Figure \ref{fig:look} was acquired on 2023 December 11, therefore we cannot guarantee that it accurately represents the coma morphology when our observations were taken (2023 December 17) {\color{black} since an outburst was reported around December 12-14}. Still it is interesting to note that the side of the coma with the most gas on our observations (South-West) coincides with with the side with the most dust on the image. {\color{black} This image indeed seems to show more dust in the West direction close to the nucleus, which then evolves into an anti-solar Northwards tail at larger distances. } \\

\paragraph*{C$_2$ profile shape:} While the CN and C$_3$ profiles still show a clear curved decrease with radial distance, the C$_2$ profiles appear almost linear. {As a sanity-check, we generated profiles of another C$_2$ band (${\Delta}v=1$, profile not shown) which showed a similar aspect.}
This particular behaviour could indicate a different origin than other species. Flatter inner-coma C$_2$ profiles have often been reported and are thought to be due to C$_2$ being a grand-daughter species or produced by icy grains in the coma \citep{combi_critical_1997}. \cite{2011Icar..213..280L} even report "C$_2$ holes" in multiple comets, where the C$_2$ density profiles dip close to the nucleus. Extended gas sources in the coma of comets have been detected in situ  \citep{1987A&A...187..801W} but still the production pathways of C$_2$ are not fully characterised. We present more evidence for extended sources in the coma of 12P in sections \ref{sec:standardhaser} and \ref{sec:bigoutburst}. \\

\paragraph*{CN feature:} Finally, on November 17th a feature is visible in the CN profile at 50000km from the nucleus, on the right side only. {This bump is visible in the spectrum taken with a longer exposure time. We therefore consider that the bump is likely real}. In section \ref{sec:minioutburst},  {we investigate whether it could be a smaller outburst following the large November 14 outburst. Such an event} was detected following the large 2023 July outburst \citep{2023ATel16194....1M,2023ATel16202....1M}. We do not detect any trace of this feature in the profile from the following night but the expansion of the gas would have likely made is fainter. We do not detect this feature in the C$_2$ and C$_3$ profiles, however the CN emissions have a higher signal to noise ratio. 

\subsubsection{Modelling by the standard Haser model} \label{sec:standardhaser}

    \begin{figure*}
        \centering
        \includegraphics[width=\linewidth]{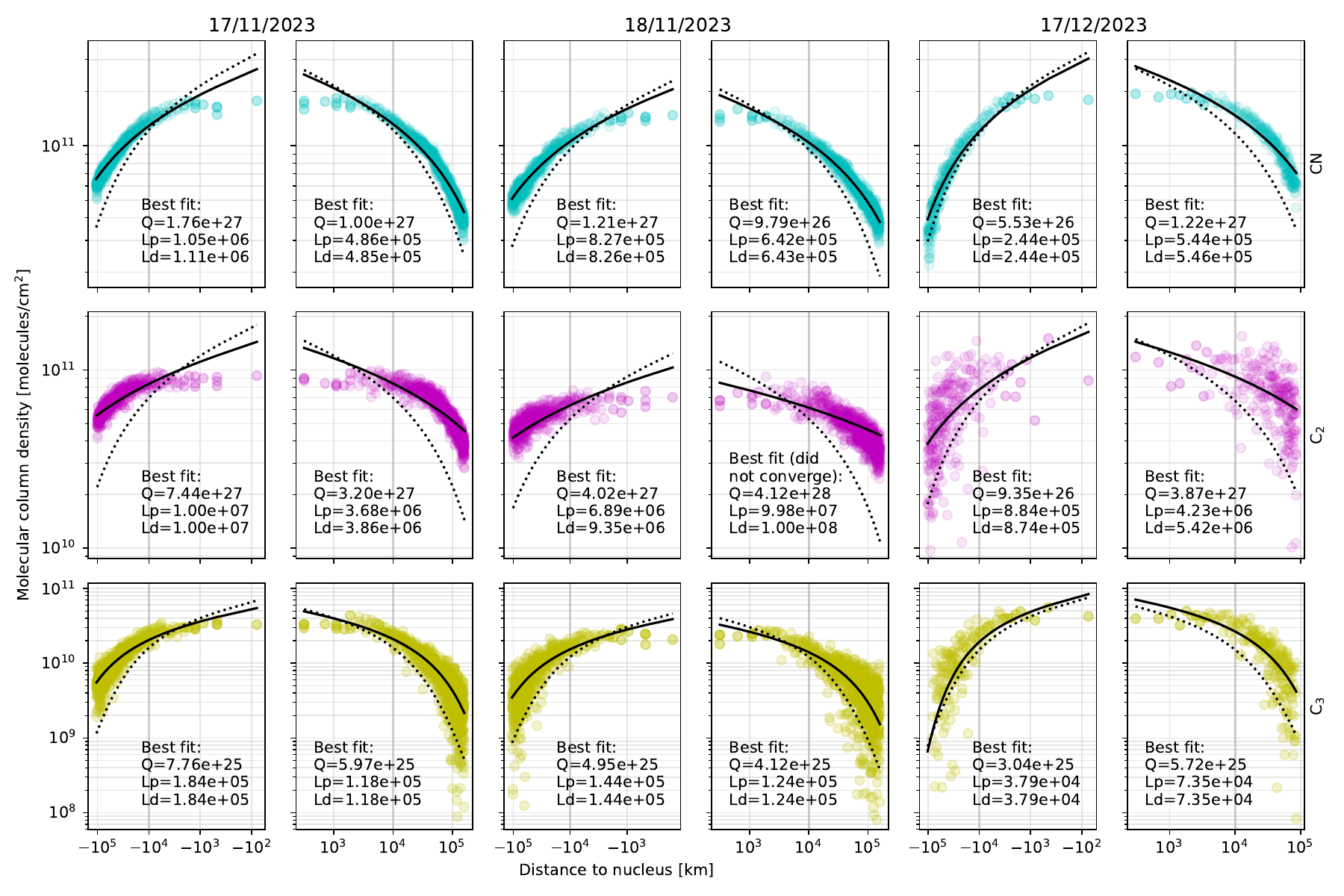}
        \caption{Molecular density profiles of CN, C$_2$ and C$_3$ in along the spectrometer's slit  in our observations from 2023 November 17 (INT-IDS), 2023 November 18 (INT-IDS) and 2023 December 17 (NOT-ALFOSC). Dotted lines show the standard Haser model profiles (eq. \ref{eq:Haser}) using parent and daughter scale-lengths from Table \ref{tab:species_parameters} and production rates adjusted to match the total flux within a 10000km aperture. Solid lines show the standard Haser model profiles with production rates and scale-lengths adjusted to match our observed profiles. The resulting parameters are listed on each subplot, with Q in molecules/second and L$_p$ and L$_d$ in kilometres.  {Both x and y axes are scaled logarithmically. Markers in the inner parts of the profiles have been made more opaque for better visibility, this does not reflect the density of data points compared to the rest of the profile.}}
        \label{fig:standardhaser}
    \end{figure*}

 {In} Figure \ref{fig:standardhaser}, we compare the observed profiles to the expected standard Haser models, i.e  given by integrating equation \ref{eq:Haser} along the line of sight using scale-lengths from Table \ref{tab:species_parameters} and our calculated production rates (Table \ref{tab:productionrates}). These theoretical profiles are represented by dotted lines. We then tried adjusting the standard Haser model with variable scale-lengths and $Q$ values to the profiles using a least-squares optimisation. The resulting profiles  {are} represented by solid lines, along with the corresponding parameters.

\paragraph*{Shapes}

Figure \ref{fig:standardhaser} shows that the expected profiles do not match the distribution of molecules in the coma at all.\footnote{ {It can be noted that} \citet{1985AJ.....90.2609C}  {proposed that the parent scale-length of C$_2$ should vary as $r_{h}^{2.5}$ rather than $r_{h}^{2}$. While this can make a significant difference at such large heliocentric distances, we have verified that the C$_2$ scale-length relationships from} \citet{1985AJ.....90.2609C}  {also do not reproduce the observed profiles.}} In this case, production rates measurements made using the method described in section \ref{subsec:productionratesmethod} strongly depend on the aperture used to measure the fluxes. This implies that the production rates that we calculated do not accurately represent the activity or composition of the comet. Therefore they should only be used for comparison with results that use the same process and parameters, such as the TRAPPIST preliminary measurements.\\

Adjusting the model to the data systematically resulted in close to equal parent and daughter scale-lengths, except for the right-side C$_2$ profile from December 18 where the algorithm tended to increasingly large scale-lengths instead of converging.
Both of these outcomes indicate that the observed profiles are too "angular" in logarithmic scale representation (flatter in the inner coma then decreasing more steeply in the outer coma) to be reproduced by the standard Haser model. 

\cite{combi_critical_1997} illustrate how the $L_{p}=L_{d}$ case translates mathematically into the most "angular" profile shape allowed by the standard Haser model. 
They show that similar or even more angular shapes can be produced by three-generation models, i.e. considering two photodissociation steps starting from grand-parent species. Such formation pathways have been proposed to explain the production of certain species and their observations, such as C$_2$H$_2$ $\rightarrow$ C$_2$H $\rightarrow$ C$_2$ \citep[e.g.][]{1997P&SS...45..721S}.
Another possible source for these radicals is that either  {they} or their parent species are produced directly from a halo of icy or CHON grains rather than from the nucleus. Extended sources have been invoked to explain the observed spatial distributions of several molecules \citep[e.g. CN, ][]{1994Icar..107..322K}. \cite{combi_critical_1997} show how their CHON halo model can also provide a better fit than the standard Haser model to some "angular" profiles measured in comet 1P/Halley. If the standard Haser model cannot reproduce our observations of the CN, C$_2$ and C$_3$ comae, it could indicate the presence of extended sources in the coma of 12P. 

For each profile, the best-fit scale-lengths differ between both sides of the coma. This is likely due to the difficulty to fit a model on a profile spanning only 100000-150000 kilometres. In particular we can see that the side of the profile that covers the shortest distance range systematically obtains longer best-fit scale-lengths. This is probably because the inner part of the profile, which is flatter, is more represented than the outer (steeper) part.
The ratio between the best-fit scale-lengths found for the left and right sides of the coma varies between species from 1.5 to 2.7 on November 17, from 1.1 to 1.3 on November 18 (omitting C$_2$ which did not converge), and 1.9 to 2.3 on December 17. While these variations are likely due to differences in data quality (different signal-to-noise between species, or a more incomplete coverage of the coma for molecules with longer lifetimes), it is interesting to see that these ratios are more similar for CN and C$_3$ on the 18th than on the 17th, as the outburst settles.
\\

\paragraph*{Production rates}
Adjusting the model to the observed profiles yields production rates that are completely different than with the basic approach. For the following analysis, let us omit the C$_2$ profile  {from 2023 November 18 } for which the fit did not converge and therefore yielded an unrealistically high production rate  {(panel at row 2 and column 4 on Figure} \ref{fig:standardhaser}). By averaging the best-fit production rates for both sides of the profiles, the resulting values are around 6 times larger for CN than what was determined in Table \ref{tab:productionrates}. For C$_3$ they are around 5 times larger, and for C$_2$ they are 27 times larger in November and 16 times larger in December.

The resulting production rates yield C$_2$/CN abundance ratios significantly larger than those obtained with theoretical scale-lengths: ranging from $169$ per cent to $422$ per cent with an average of $312$ per cent. 
As our adjusted models still do not perfectly match the observed profiles, especially for C$_2$, these adjusted production rates may not accurately represent the composition of 12P, but this discrepancy between measurement methods highlights how challenging composition studies can be for comets with atypical behaviours and what biases may arise in large standardised surveys. It could also explain the variations seen in the TRAPPIST abundance ratios (Figure \ref{fig:trappist}), in particular if the amount of extended sources varies overtime or is affected by the sudden outbursts.\\

The next section will focus on the observations from November and investigate whether this departure from the standard Haser model can be explained by the outburst that happened on November 14, as the production rate varying throughout the outburst can give the coma profiles a different shape.

\subsubsection{Modelling of the large November 14 outburst} \label{sec:bigoutburst}

\begin{figure*}
    \centering
    \includegraphics[width=\linewidth]{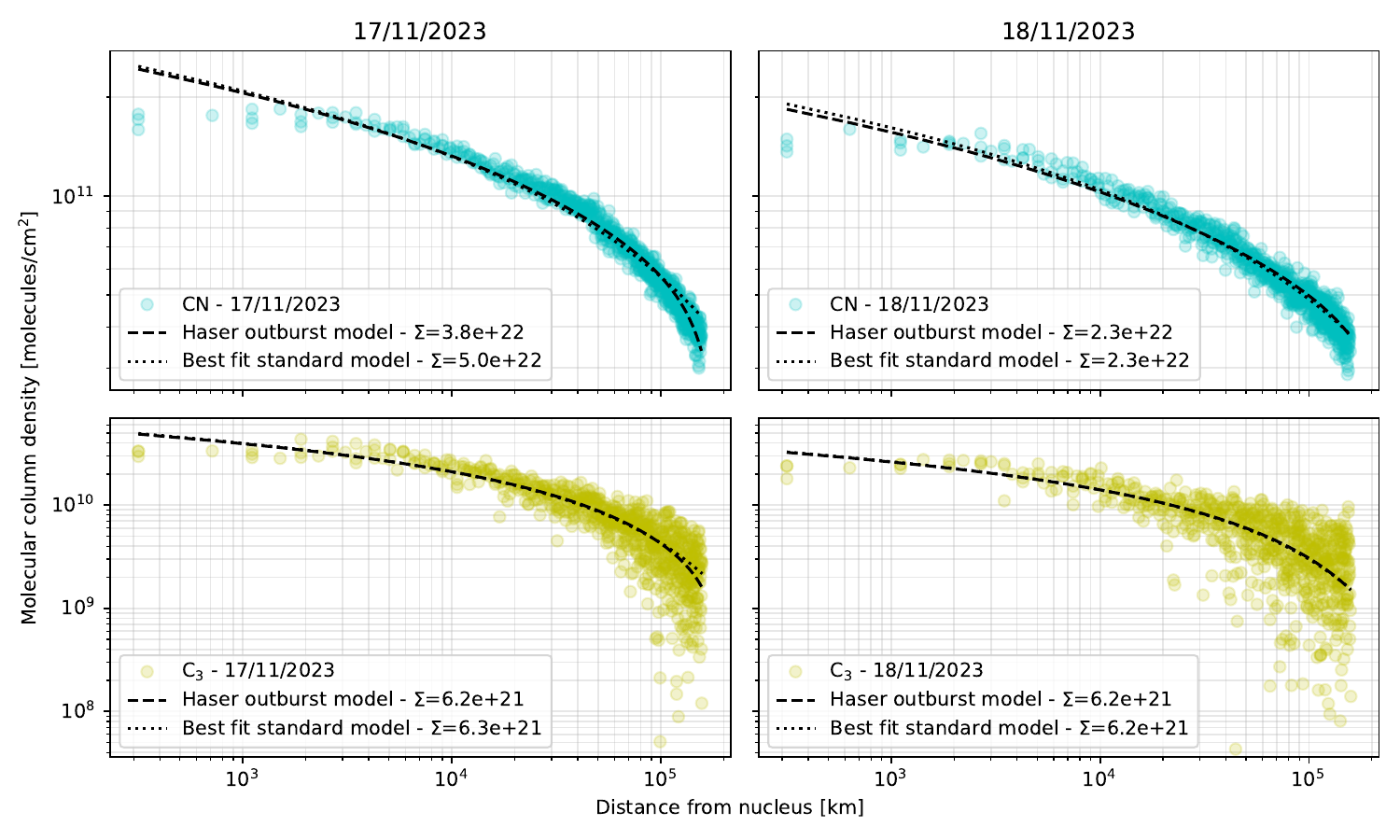}
    \includegraphics[width=\linewidth]{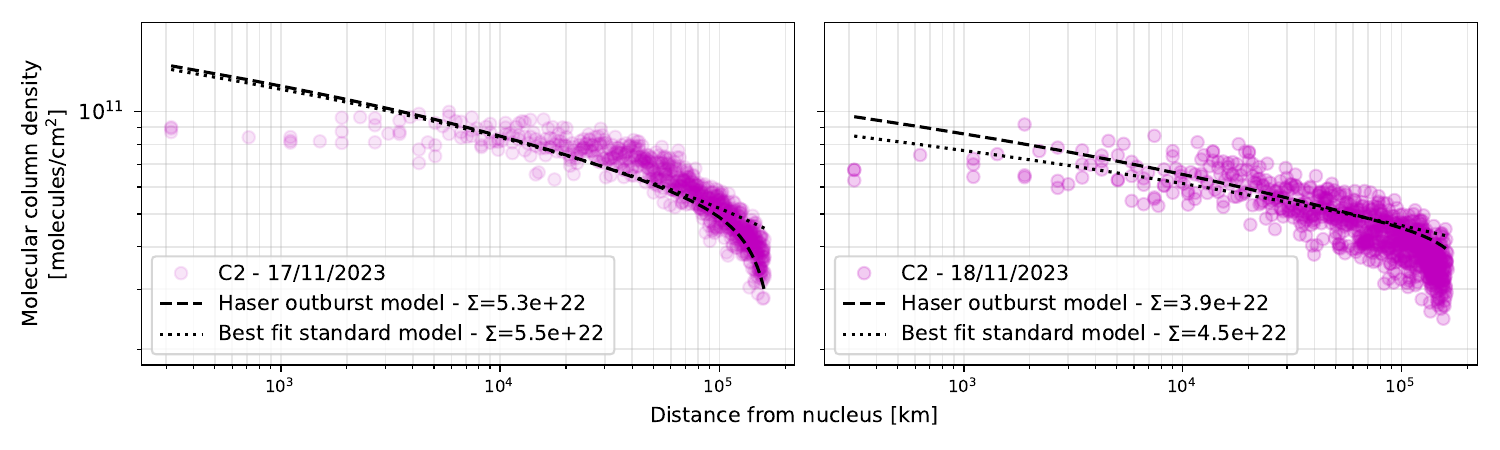}
    \caption{ {Comparison of the best-fit standard Haser model (dotted) and best fit outburst Haser model (dashed). In the legends $\Sigma$ is the residual sum of squares, minimised by the fitting algorithm.  } \textbf{Top and middle rows:} Output of simultaneous least-square fitting the outburst model (eq \ref{eq:Haseroutburst1} and \ref{eq:Haseroutburst2}) to the CN and C$_3$ profiles from November 17 and November 18. 
    Here the model parameters are Q$_{0}$(CN)=$1.5\times10^{26}$~s$^{-1}$, Q$_{1}$(CN)=$1.3\times10^{27}$~s$^{-1}$, L$_{p}$(CN)=$2.4\times10^{5}$km, L$_{d}$(CN)=$3.1\times10^{5}$km, Q$_{0}$(C$_{3}$)=$5.7\times10^{24}$~s$^{-1}$, Q$_{1}$(C$_{3}$)=$1.2\times10^{26}$~s$^{-1}$, L$_{p}$(C$_{3}$)=$7.3\times10^{4}$km, L$_{d}$(C$_{3}$)=$7.3\times10^{4}$km, v$_{1}=0.73$km~s$^{-1}$, r$_{a}$/v$_{1}=1.3\times10^{4}$s, r$_{b}$/v$_{1}=1.7\times10^{5}$s, and an outburst peak on 2023 November 15, 01:54UT. 
    \textbf{Bottom row:} Output of least-square fitting of the outburst model C$_2$ profiles from November 17 and November 18 simultaneously. Values of v$_{1}$, r$_{a}$, r$_{b}$ and the outburst peak time were fixed to the ones found from adjusting the model to the CN and C$_3$ profiles, resulting in Q$_{0}$(C$_{2}$)=$1.0\times10^{26}$s$^{-1}$,  Q$_{1}$(C$_{2}$)=$2.9\times10^{28}$s$^{-1}$, L$_{p}$(C$_{2}$)=$8.1\times10^{6}$km and L$_{d}$(C$_{2}$)=$8.5\times10^{6}$km.
    }
    \label{fig:haseroutburst}
\end{figure*}

From the decrease in production rates that we observe between November 17 and November 18 it is clear that our profiles still represent the aftermath of the large November 14 outburst, which is bound to affect the coma shape. Indeed, the amount of molecules at a given distance from the nucleus depends on the production rate at the time of ejection. In the case of a post-outburst profile, the production has previously varied through time, as opposed to the constant production rate assumed by the steady-state Haser model. In this section we aim to explore whether this effect alone can explain why the standard Haser model does not fit the data. 

\paragraph*{CN and C$_3$ fits:} We attempted to adjust the outburst model described in section \ref{subsec:radialprofilesmethod} to the CN and C$_3$ profiles from November 17 and November 18 simultaneously, in an attempt to  {constrain} values of Q$_{0}$, Q$_{1}$, L$_{p}$, L$_{d}$, v$_{1}$, r$_{a}$, r$_{b}$ and the outburst peak time. As demonstrated by \cite{2016A&A...589A...8O}, this model can reproduce bumps in radial profiles that are sometimes observed during outbursts. We only considered the right-hand side of the profiles as they cover larger nucleocentric distances. We did not attempt to include the C$_2$ profiles as their shape initially seemed too different from the standard model. Assuming a steady-state expansion velocity of 1km~s$^{-1}$, most of the material released at the peak of the outburst perpendicularly to the line of sight has already left the field of view by November 17, meaning that the data contain less information about the timescales of the outburst, making it hard to constrain some of the parameters. Still, this analysis can show whether the outburst model can generate Haser profiles that better match our observations.
Initial conditions were chosen based on the theoretical scale-lengths from Table \ref{tab:species_parameters}, expected orders of magnitudes of the production rates, observed timescales of the outburst\footnote{Nick James, \href{https://britastro.org/section_news_item/comet-12p-pons-brooks-outburst-continue}{British Astronomical Association}}  and a velocity of 1km~s$^{-1}$. The outcome of the minimisation algorithm  {and the resulting residuals are} presented in Figure \ref{fig:haseroutburst}.
{We can see that, at the distances covered by our profiles, the outburst model looks quite similar to the standard Haser model. For CN on the first night, the outburst model seems to better reproduce the decrease of the profile past 100000km. On the second night the two models are nearly  indistinguishable.} 
For C$_3$ the difference between the two models is very minor and would mostly be visible in regions where the observed signal is very faint, therefore both models provide an equally good fit.

The CN production rates  (pre-outburst Q$_{0}$(CN)=$1.5\times10^{26}$s$^{-1}$ {with an additional outburst source
of CN that was} Q$_{1}$(CN)=$1.3\times10^{27}$s$^{-1}$) seem consistent with the comet's activity, as Q$_{0}$(CN) is of the same order of magnitude as \citep{jehin_trappist_2023-2} measured on November 12.
The best fit C$_3$ production rates are Q$_{0}$(C$_{3}$)=$5.7\times10^{24}$s$^{-1}$ and Q$_{1}$(C$_{3}$)=$1.2\times10^{26}$s$^{-1}$. Q$_{0}$(C$_{3}$) is a factor 2 lower than what \cite{jehin_trappist_2023-2} measured pre-outburst, but  {we} deem this to be an acceptable order of magnitude, especially since the resulting C$_3$/CN ratio Q$_{0}$(C$_{3}$)/Q$_{0}$(CN)=3.8 per cent is typical.  For both species, the resulting parent and daughter scale-lengths are still almost equal, with  L$_{p}$(CN)=$2.4\times10^{5}$km, L$_{d}$(CN)=$3.1\times10^{5}$km,  L$_{p}$(C$_{3}$)=$7.3\times10^{4}$km and L$_{d}$(C$_{3}$)=$7.3\times10^{4}$km. This indicates that the nature of the activity at the time (outburst rather than steady-state) does not explain why the profiles depart from a Haser profile, and the presence of extended sources is still likely. 
The outburst material velocity of v$_{1}=0.73$km~s$^{-1}$ is lower than the velocity that we assumed for the Haser model but acceptable at this heliocentric distance ($r_{h}\approx3.5$au), especially if gas is being released from slower grains. The outburst peak time of November 15 01:54 UT is compatible with observations by multiple astronomers monitoring the comet\footnotemark[4]. Characteristic timescales of the outburst r$_{a}$/v$_{1}=1.3\times10^{4}$s and r$_{b}$/v$_{1}=1.7\times10^{5}$s appear realistic since we observe a return to steady state within a few days and the outburst was reported to take off within a few hours only\footnotemark[4]. These are also similar to the typical timescales observed for several large outbursts of comet 29P \citep{2008A&A...485..599T}, which also undergoes frequent outbursts.
We attempted to allow different velocities for CN and C$_3$ instead of a common v$_1$ but this lead to similar results overall and velocities close to equal for both species. 

\paragraph*{C$_2$ profile:} We then attempted to model the C$_2$ profiles from November 17 and 18 simultaneously to determine Q$_{0}$(C$_{2}$), Q$_{1}$(C$_{2}$), L$_{p}$, L$_{d}$, but using the values of v$_{1}$, r$_{a}$, r$_{b}$ and the outburst peak time determined from CN and C$_3$. Considering the strange aspect of the C$_2$ profile and the data quality, we preferred this approach to trying to fit all parameters at the same time. Initially, the model converged towards Q$_{0}$(C$_{2}$) values that are significantly lower than expected ($\approx1\times10^{19}$s$^{-1}$). However, imposing a more realistic order of magnitude Q$_{0}$(C$_{2}$)$>1\times10^{26}$s$^{-1}$ resulted in adjusted profiles that are visually just as satisfactory as letting Q$_{0}$(C$_{2}$) vary freely. In this case we obtained Q$_{0}$(C$_{2}$)=$1.0\times10^{26}$s$^{-1}$,  Q$_{1}$(C$_{2}$)=$2.9\times10^{28}$s$^{-1}$, L$_{p}$(C$_{2}$)=$8.1\times10^{6}$km and L$_{d}$(C$_{2}$)=$8.5\times10^{6}$km. The corresponding profile is shown on Figure \ref{fig:haseroutburst}. 
 Although we could not accurately constrain Q$_{0}$(C$_{2}$), nor Q$_{1}$(C$_{2}$) which seems abnormally high compared to Q$_{1}$(CN), we show that the outburst model can match the aspect of the observed C$_2$ profiles slightly better than the standard model {at large distances}. However, visually it still does not provide as good of a fit as for CN and C$_3$, and as for these species equal scale-lengths are required.

 {We conclude that the outburst does not completely explain why the standard (non-outburst) Haser model does not apply, as the outburst model requires non-physical scale-lengths for all molecules and still to match the C$_2$ profile shape.
We propose} that extended sources must contribute to the production of all species, but that extended sources of a different nature or reaction chains more complex than parent/daughter might have to be considered for C$_2$ as well.

\begin{figure}
    \centering
    \includegraphics[width=\linewidth]{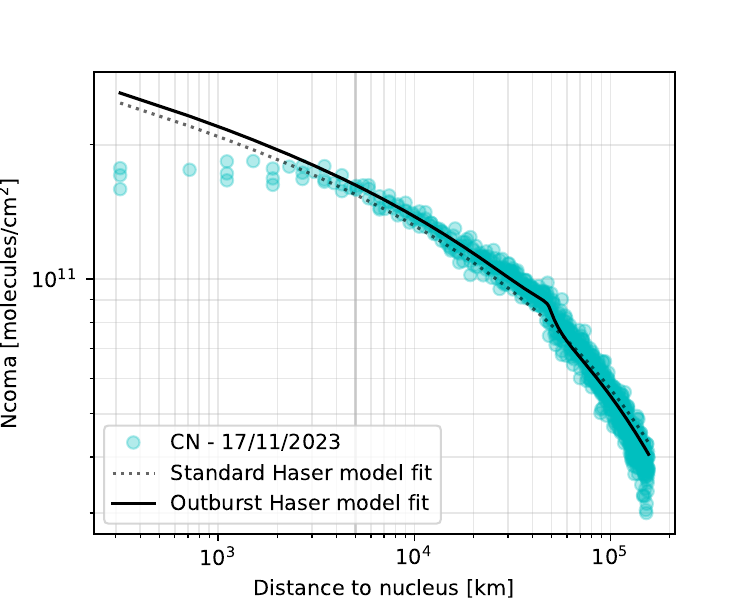}
    \caption{CN profile from 2023 November 17 along with the best fit for the standard Haser model (dotted line) and for the outburst model (solid line) adjusted to reproduce the small feature visible at $\sim$50000km. Because the standard Haser model is particularly far from the observations close to the nucleus, the outburst model was adjusted on the data at distances greater than 5000km. A grey vertical line indicates this cut-off.}
    \label{fig:minioutburst}
\end{figure}

\subsubsection{Modelling of a possible November 17 mini-outburst} \label{sec:minioutburst}

Finally, we tried adjusting the outburst model to the feature visible in the CN profile from November 17 (Figure \ref{fig:profiles}).   {If this is an outburst,} because we do not have information on the temporal evolution of this feature we imposed an arbitrary outburst material expansion velocity of v$_{1}$=$1$km~s$^{-1}$.  {However depending on the nature of the outburst, using the same velocity as for the steady-state might be erroneous.} Because the standard Haser model does not correctly reproduce the overall profile to begin with (especially at short distances) and in order to best model the shape of the outburst feature over the steady-state baseline, we only adjusted this model to the profile past 5000km. The resulting profile is shown on Figure \ref{fig:minioutburst}, corresponding to the following parameters: 
L$_{p}$(CN)=$3.0\times10^{5}$km, L$_{d}$(CN)=$6.2\times10^{5}$km,
Q$_{0}$(CN)=$6.48\times10^{26}$s$^{-1}$ and Q$_{1}$(CN)=$3.8\times10^{26}$s$^{-1}$,
r$_{a}$/v$_{1}=5.1\times10^{3}$s, r$_{b}$/v$_{1}=5.3\times10^{3}$s, ${\Delta}t$=$4.8\times10^{4}$s. 
Because the flattest part of the profile was masked, we obtain parent and daughter scale-lengths that are not equal but still of similar orders of magnitude, which would not be the case of the expected values. 

 {This shows that the outburst model can produce features similar to the one that we see in our CN profile. The corresponding event could then be a short outburst with a production rate of one order of magnitude below the total peak outburst production rate found in section} \ref{sec:bigoutburst}.  {However, with our data alone, we cannot guarantee that this feature is real. }
We do not find any strong evidence for a dust-counterpart to this event in the LCO data.

\section{Conclusion} \label{sec:conclusion}

In this paper, we analysed long-slit optical spectra of comet 12P/Pons-Brooks acquired between 2023 August and 2023 December, quantifying the comet's composition in daughter species and comparing the molecular density profiles along the slit to multiple models. In particular, we hoped that spectra obtained on consecutive nights soon after the large 2023 November 14 outburst could provide insight about the nature and behaviour of the outburst.

Assuming that the distribution of molecules in the coma follows the standard Haser model with commonly used parent and daughter scale-lengths, our measured production rates show a "typical" composition with a C$_2$/CN ratio of about 90 per cent, which does not seem to change throughout the outburst or from November to December, and is in agreement with measurements by other teams around that time.

However, calculated coma profiles of CN, C$_2$ and C$_3$ indicate that the behaviour of the gas coma is more complex, with asymmetries and separate species behaving differently.
Comparing our profiles to the standard Haser model computed with empirical parent and daughter scale-lengths shows that it does not match the observed coma shape. This result invalidates the composition measurements that were made under the assumption that this model was a valid representation of the coma.

After adjusting the standard Haser model to the data, the model still does not provide a good match and the best fit scale-lengths are equal for parent and daughter species, indicating that a more complex model (e.g. icy grains, CHON grains, or three generations) might be necessary. Best fit production rates result into a larger C$_2$/CN ratio, which highlights how inaccurate scale-lengths or models can introduce bias in composition measurements.

Comparing the profiles from November to an outburst model, which accounts for the variation through time of the production rate during the outburst, we showed that this model reproduces the shape of the profiles better than the steady-state model for CN and C$_3$, although equal parent and daughter scale-lengths are still required. However this model fails to match the shape of the C$_2$ coma. We propose that extended sources contribute to the production of all species in the coma, and that the C$_2$ coma is particularly affected by different types of extended sources and/or complex formation pathways.

Finally we showed that it is possible that a small short outburst happened between the large 2023 November 14 outburst and our observations, which would explain a small feature on our November 17 CN profiles.\\

{\color{black} 
12P is yet another comet for which simple models cannot reproduce large-scale gas distributions, in particular for C$_2$. More needs to be known about the production mechanisms of these species in comae to understand how these observed distributions reflect the "true" composition of the ice. 
Radio or infrared observations of parent species in 12P's coma could provide insight into the formation mechanisms of the daughter species that are observed in the optical range. 
A more in depth analysis of 12P's gas and dust production through time and through outbursts could also help characterise and clarify the origin of 12P's variable activity.} {\color{black} 
In this context, we hope that other observing campaigns can improve our understanding of the nature of 12P's activity.
}

\section*{Acknowledgements}

The authors would like to thanks U. G. Jørgensen and C. Snodgrass for the opportunity to obtain NOT observations.

R. Kokotenekova would like to acknowledge support from “L’Oreal UNESCO For Women in Science” National program for Bulgaria.

This work makes use of observations from the 1m telescopes and Sinistro instruments of the Las Cumbres Observatory global telescope network. Data were obtained under the LOOK Proposal (Proposal code LTP2023B-001) and raw and reduced data are available from the LCO Science Archive at \href{https://archive.lco.global}{https://archive.lco.global}.

This paper is based on observations made with the MuSCAT3 instrument, developed by the Astrobiology Center and under financial supports by JSPS KAKENHI (JP18H05439) and JST PRESTO (JPMJPR1775), at Faulkes Telescope North on Maui, HI, operated by the Las Cumbres Observatory.

The Comet Chasers schools outreach project is funded by the UK Science and Technology Facilities Council through the DeepSpace2DeepImpact Project, Society, the Open University and Cardiff University.  It accesses the LCOGT telescopes through the Faulkes Telescope Project (FTPEPO2014A-004), which is partly funded by the Dill Faulkes Educational Trust. Observations include those made by Cai Stoddard-Jones and students from St Marys Catholic Primary School, Bridgend, Wales.

 {For the purpose of open access, the author has applied a Creative Commons Attribution (CC BY) licence to any Author Accepted Manuscript version arising from this submission.}

 {The authors would like to thank the referee for their useful insight to improve this manuscript.}


\section*{Data Availability}
 
The data underlying this article will be shared on reasonable request to the corresponding author.



\bibliographystyle{mnras}
\bibliography{example}





\bsp	
\label{lastpage}
\end{document}